\documentclass[%
reprint,
superscriptaddress,
 amsmath,amssymb,
 aps,
prl,
]{revtex4-1}
\usepackage{graphicx,mathrsfs}
\usepackage{dcolumn}
\usepackage{bm}
\usepackage[caption=false]{subfig}

\usepackage[colorinlistoftodos]{todonotes}

\begin{document}

\preprint{APS/123-QED}

\title{Observation and uses of position-space Bloch oscillations in an ultracold gas}

\author{Zachary A. Geiger}
\thanks{Equal contributions.}
\affiliation{University of California and California Institute for Quantum Emulation, Santa Barbara CA 93105}

\author{Kurt M. Fujiwara}
\thanks{Equal contributions.}
\affiliation{University of California and California Institute for Quantum Emulation, Santa Barbara CA 93105}

\author{Kevin Singh}
\affiliation{University of California and California Institute for Quantum Emulation, Santa Barbara CA 93105}

\author{Ruwan Senaratne}
\affiliation{University of California and California Institute for Quantum Emulation, Santa Barbara CA 93105}

\author{Shankari V. Rajagopal}
\affiliation{University of California and California Institute for Quantum Emulation, Santa Barbara CA 93105}

\author{Mikhail Lipatov}
\affiliation{University of California and California Institute for Quantum Emulation, Santa Barbara CA 93105}

\author{Toshihiko Shimasaki}
\affiliation{University of California and California Institute for Quantum Emulation, Santa Barbara CA 93105}

\author{Rodislav Driben}
\affiliation{Department of Physics and CeOPP, University of Paderborn, D-33098, Paderborn, Germany}

\author{Vladimir V. Konotop}
\affiliation{Centro de F\'isica Te\'orica e Computacional and Departamento de F\'isica, Faculdade de Ci\^encias, Universidade de
Lisboa, Campo Grande, Ed. C8, Lisboa 1749-016, Portugal}

\author{Torsten Meier}
\affiliation{Department of Physics and CeOPP, University of Paderborn, D-33098, Paderborn, Germany}

\author{David M. Weld}
\email[Corresponding author: ]{weld@ucsb.edu}
\affiliation{University of California and California Institute for Quantum Emulation, Santa Barbara CA 93105}

\begin{abstract}
We report the direct observation and characterization of position-space Bloch oscillations using an ultracold gas in a tilted optical lattice.  While Bloch oscillations in momentum space are a common feature of optical lattice experiments, the real-space center-of-mass dynamics are typically too small to resolve. Tuning into the regime of rapid tunneling and weak force, we observe real-space Bloch oscillation amplitudes of hundreds of lattice sites, in both ground and excited bands. We demonstrate two unique capabilities enabled by tracking of Bloch dynamics in position space: measurement of the full position-momentum phase-space evolution during a Bloch cycle, and direct imaging of the lattice band structure. These techniques, along with the ability to exert long-distance coherent control of quantum gases without modulation, may open up new possibilities for quantum control and metrology.  
\end{abstract}

\maketitle

Quantum particles in a periodic potential exhibit an oscillatory response to constant forces~\cite{Bloch1929,Zener523}.  During these Bloch oscillations, both the quasimomentum and the position of the particles evolve periodically, as a direct observable consequence of the periodicity of the band structure. The resultant localization is a fundamental feature of coherent transport in a lattice. In conventional electronic systems rapid decoherence complicates the observation of Bloch oscillations, though their realization is possible using superlattices~\cite{meier-BOsuperlattice} and photonic waveguide lattices~\cite{opticalblochosc}. Ultracold atomic gases in optical lattices, however, provide a nearly ideal platform for the observation of momentum-space Bloch oscillations~\cite{dahan1996,kasevichatomintf} and related phenomena, including interaction-dependent effects~\cite{arimondo-meanfieldBOs,inguscio-interactionBOs,nagerlinteractionBOs,Gaul2011,Meinert2014}, spatial breathing modes \cite{Preiss2015}, and control of Bloch dynamics using applied modulations~\cite{Haller2010,Alberti2009}. However, even in cold atom experiments, direct observation of position-space Bloch oscillations of the center of mass has remained elusive. Position-space center-of-mass Bloch oscillations have generally either been inferred through observations of the quasimomentum evolution~\cite{dahan1996}, or artificially magnified using dynamical perturbations for rectification of Bloch dynamics~\cite{Haller2010}.  Thus, in a sense, Zener's original conception of Bloch oscillations has yet to be directly observed with cold atoms.

\begin{figure}[t!]
\includegraphics[width=\linewidth]{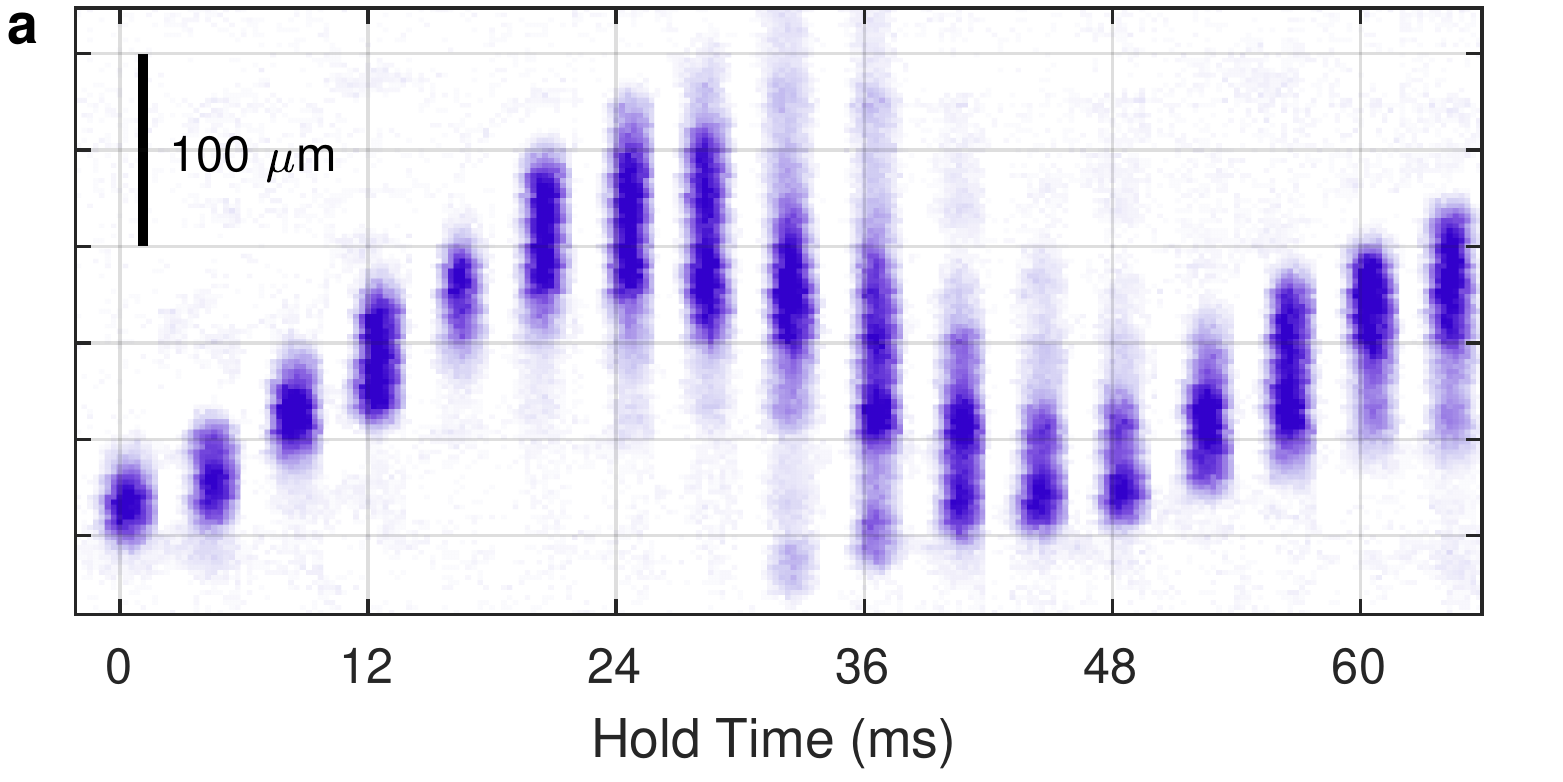}
\includegraphics[width=\linewidth]{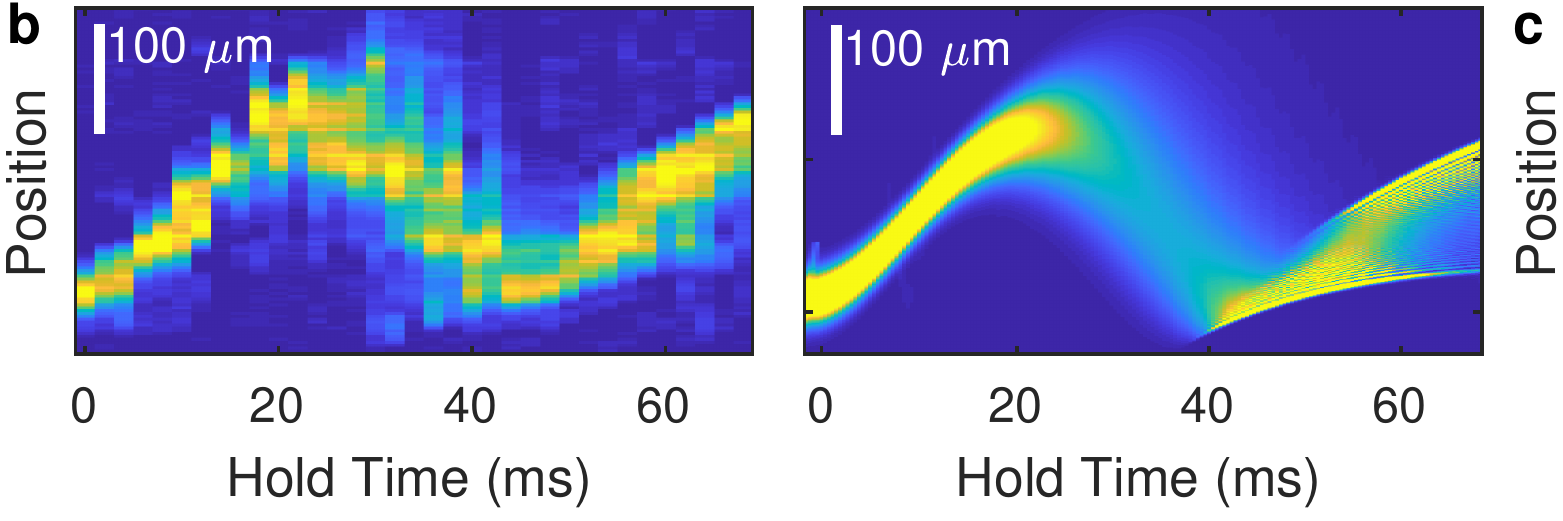}
\caption{Position-space Bloch oscillations. \textbf{(a)} Time sequence of in-situ absorption images of an atomic ensemble in a 5.4~$E_R$-deep optical lattice subjected to a force corresponding to a Bloch frequency of 21.1~Hz. Each image corresponds to an individual experimental run with the indicated hold time. \textbf{(b)} Measured evolution of atomic density distributions in the lattice direction during a Bloch oscillation. \textbf{(c)} Numerical GPE prediction for (b), using the split-step Fourier method. Asymmetrical width variation is due to force inhomogeneity.}
\label{fig:mixOD}
\end{figure}
In this letter, we report the observation and characterization of position space-Bloch oscillations using ultracold $^7$Li in a tilted optical lattice. Lithium's low mass facilitates the simultaneous realization of fast tunneling and weak tilt which are required for observation of real-space Bloch oscillations, and the shallow zero crossing of lithium's Feshbach resonance allows elimination of interactions. Our experiments begin with a Bose condensate of lithium in a single-beam optical lattice. The mean-field behavior can in general be described by the Gross-Pitaevskii  equation (GPE) $i\hbar\frac{\partial\Psi}{\partial t}=-\frac{\hbar^2}{2m}\nabla^2\Psi + V(r,z)\Psi+\mathcal{F}z\Psi+g \left|\Psi\right|^2\Psi.
$
 Here $V=-V_{L}\cos^2\left({\pi z/d}\right)\exp\left(-{r^2}/{2\sigma^2}\right)+m\omega_{0}^2z^2/2$ is the potential in cylindrical coordinates due to the lattice beams with wavelength $\lambda\! =\! 2d\! =\! 1064$~nm and a parabolic trap of frequency $\omega_0\!\simeq\! 2\pi\times 15.6$~Hz in the $z$ direction, $\sigma\!\simeq\! 42.5\ \mu$m is the transverse trap width, $\mathcal{F}$ is the applied force, $m$ is the atomic mass, and $g$ is the interaction amplitude which we make negligible by Feshbach tuning. This 3D potential is non-separable, but as transverse dynamics do not play a significant role a 1D Bose-Hubbard Hamiltonian, 
\begin{equation}
\hat{H}= - J \sum_{i} \hat{a}_{i+1}^{\dagger}\hat{a}_{i} + \sum_{i}\frac{U}{2}\hat{n}_{i}(\hat{n}_{i}-1)+ F \sum_{i} i \hat{n}_i, \label{eq1}
\end{equation}
provides an alternative tight-binding single-band description capturing the main features of Bloch oscillations.
In Eq.~(\ref{eq1})  $\hat{a}_{i}^{\dagger}$ and $\hat{a}_{i}$ are bosonic creation and annhilation operators at lattice site $i$,  $J$ is the tunneling energy, $\hat{n}_{i}=\hat{a}_{i}^{\dagger}\hat{a}_{i}$, $U$ is the onsite interaction energy, and $F=\mathcal{F}d$ the energy offset per lattice site due to the force $\mathcal{F}$. Below we quote values for $F$ and $J$ in Hz; multiplication by Planck's constant $h$ yields an energy in Joules. By Feshbach tuning to the scattering length zero-crossing we are able to operate in the $J\gg F \gg U$ regime for the entirety of the experiment, avoiding interaction-induced dephasing and instabilities~\cite{buchleitner2003,VVKgapsolitons,VVK-longBOs,Driben-nonlinBOs}. When an atomic ensemble is subjected to this Hamiltonian in this regime, the center of mass position oscillates at the Bloch frequency $f_B=F/h$ with oscillation amplitude given by the Wannier-Stark localization length $l_{WS}=2J/F$~\cite{Zener523,dahan1996}.

The experiments begin with creation of a Bose condensate of approximately $10^5$ $^7$Li atoms in the $\left | F=1,m_F=1\right\rangle$ hyperfine state in a crossed optical dipole trap.  After evaporation, the magnetic field is ramped to the scattering length zero crossing near 543.6~G in 100~ms~\cite{Pollack2009}.  The condensate is then loaded into the optical lattice in 100~ms, initializing the atomic ensemble into the ground band around zero quasimomentum. The lattice depth $V_L$ is adjusted between 4 and 15~$E_R$ (calibrated with amplitude modulation spectroscopy), where $E_R=\hbar^2 k_L^2/2 m$ is the recoil energy and $k_L=2\pi/\lambda$ is the lattice wavevector. For a given $V_L$, band structure calculations determine $J$.  A magnetic field gradient along the lattice direction creates a tunable applied force giving rise to a Bloch frequency $f_B$ between 20 and 50~Hz.  The optical dipole trap is suddenly switched off to initiate Bloch oscillations.  For an atomic ensemble in a band with a dispersion relation $E(k)$, the mean position $x(t)$ evolves in time according to the group velocity: 
\begin{equation}
\frac{d x(t)}{d t}=\frac{1}{\hbar}\frac{\partial E(t)}{\partial k}.
\label{eq:com}
\end{equation}After a variable hold time during which Bloch oscillations occur, the in-situ spatial distribution of the atomic ensemble is measured using absorption imaging. 
\begin{figure}[t]
\includegraphics[width=\linewidth]{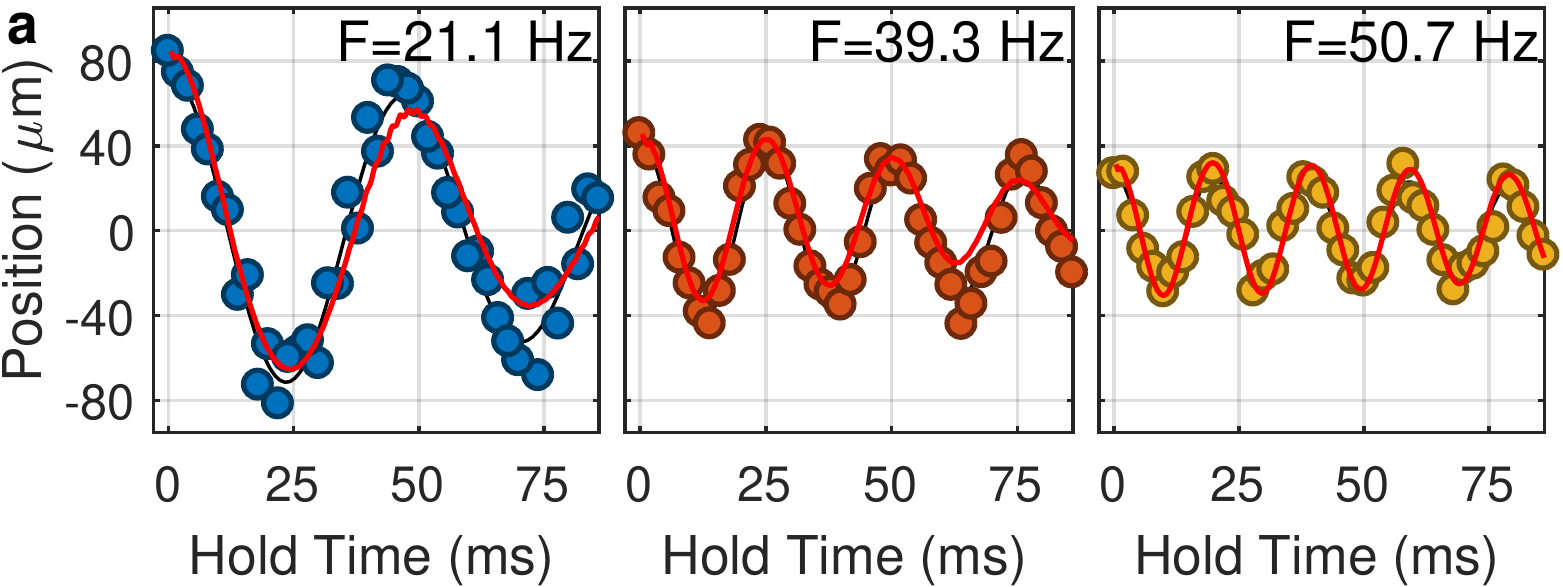}
\includegraphics[width=\linewidth]{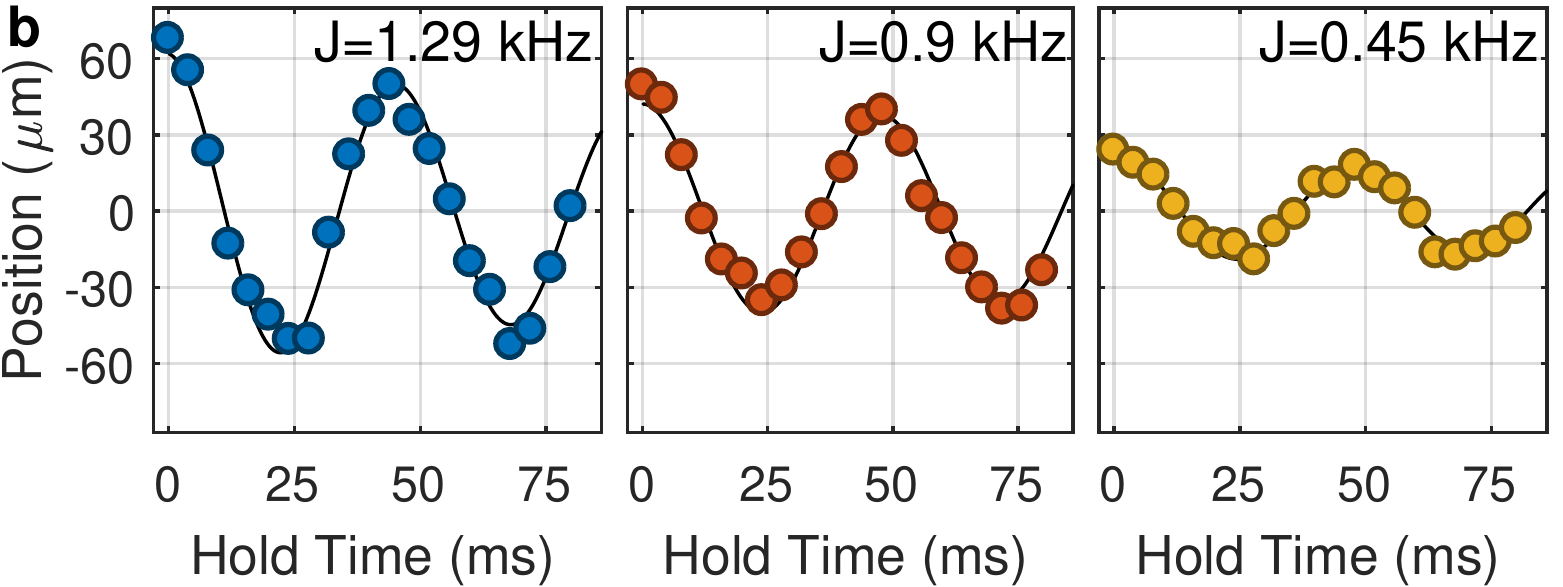}
\subfloat{\includegraphics[width =.5\linewidth]{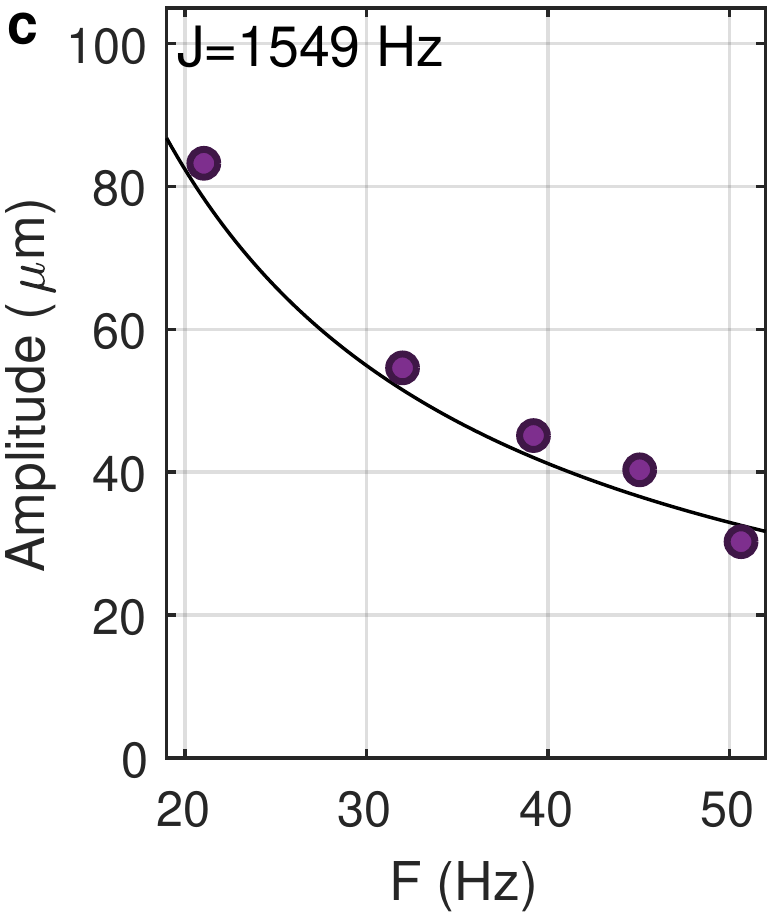}} 
\subfloat{\includegraphics[width =.5\linewidth]{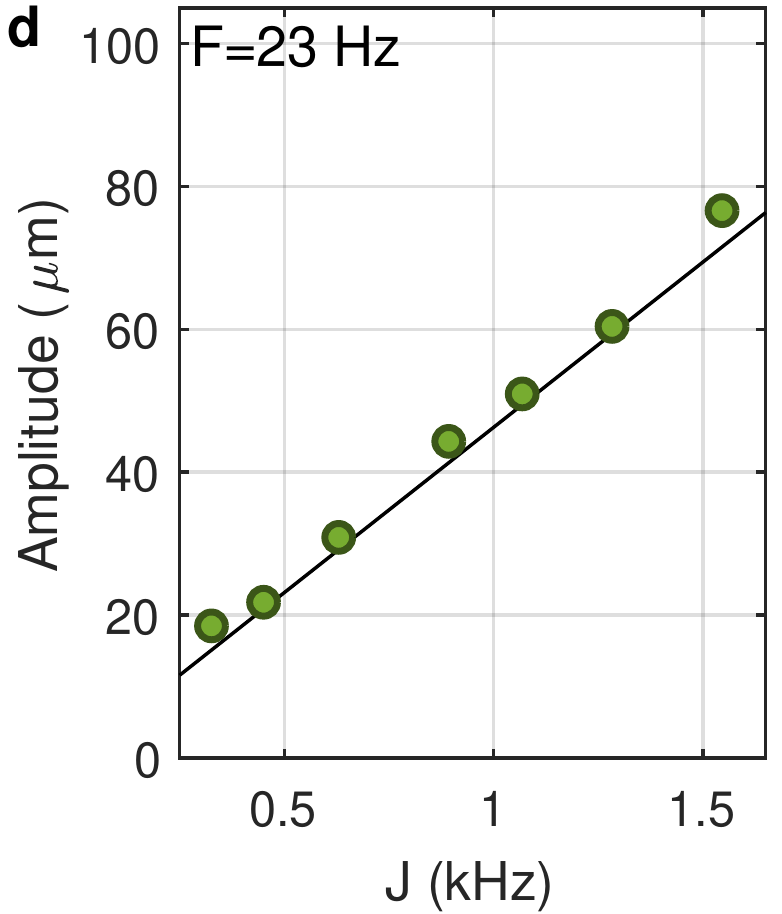}}\\
\caption{Characterization of position-space Bloch oscillations. \textbf{(a)} Ensemble position versus time during Bloch oscillations at different forces with a constant tunneling rate $J=1549$~Hz.  \textbf{(b)} Ensemble position versus time during Bloch oscillations with different tunneling rates at a constant force of $F=23$~Hz. Solid black lines in (a) and (b) are fits to damped sinusoids while solid red lines are the result of numerical GPE calculations. \textbf{(c)} Bloch oscillation amplitude as a function of force at $J=1549$~Hz. \textbf{(d)} Bloch oscillation amplitude as a function of tunneling rate at $F=23$~Hz.  Solid lines in (c) and (d) represent the predicted $l_{WS}=2J/F$ with no fit parameters. }
\label{fig:BOvsForce}
\end{figure}

Fig.~\ref{fig:mixOD} presents a typical measurement of position-space Bloch oscillations.  As the quasimomentum linearly increases, the center of mass position evolves according to Eq.~(\ref{eq:com}), changing direction after a Bragg scattering event at the edge of the Brillouin zone and oscillating with an amplitude of over 150~$\mu$m ($\simeq$300 lattice sites). The measured oscillation amplitude and frequency are consistent with theoretical predictions for $l_{WS}$ and $f_B$, and the detailed dynamics are in agreement with the results of numerical GPE calculations shown in Fig.~\ref{fig:mixOD}c. The direct measurement of position-space atomic Bloch oscillations of the center of mass in the absence of any external modulation is the first main result of this paper.

To characterize the position-space Bloch dynamics, we directly probe the dependence of the Wannier-Stark localization length $l_{WS}$ on tunable system parameters by repeating the measurement of Fig.~\ref{fig:mixOD} at varying values of $J$ and $F$.  Fig.~\ref{fig:BOvsForce}a shows position-space Bloch oscillations for different applied forces at a constant tunneling rate $J=1549$~Hz. The measured initial oscillation amplitude is inversely proportional to the force (Fig.~\ref{fig:BOvsForce}c), and agrees quantitatively with the expected localization length $l_{WS}$.  The observed decrease of the oscillation amplitude with increasing hold time is consistent with expectations due to the residual harmonic confinement, and with numerical GPE calculations. Fig.~\ref{fig:BOvsForce}b and Fig.~\ref{fig:BOvsForce}d present a complementary measurement, in which tunneling rates were varied at a constant force.  The measured oscillation amplitude increases linearly with the tunneling rate $J$ and is again in agreement with the theoretically expected value of $l_{WS}$, with no adjustable fit parameters. 

\begin{figure}[t!]
\includegraphics[width=\linewidth]{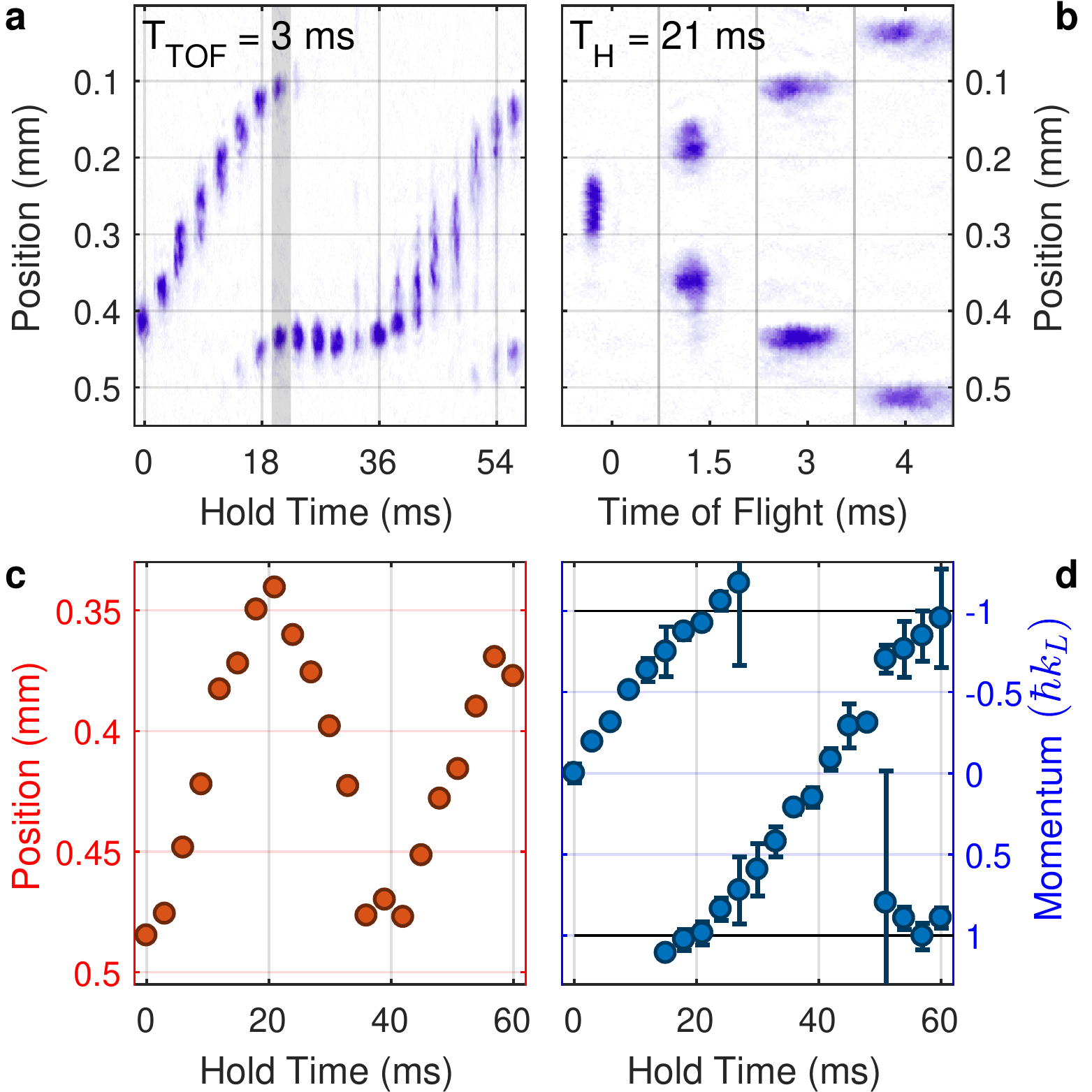}
\includegraphics[width=\linewidth]{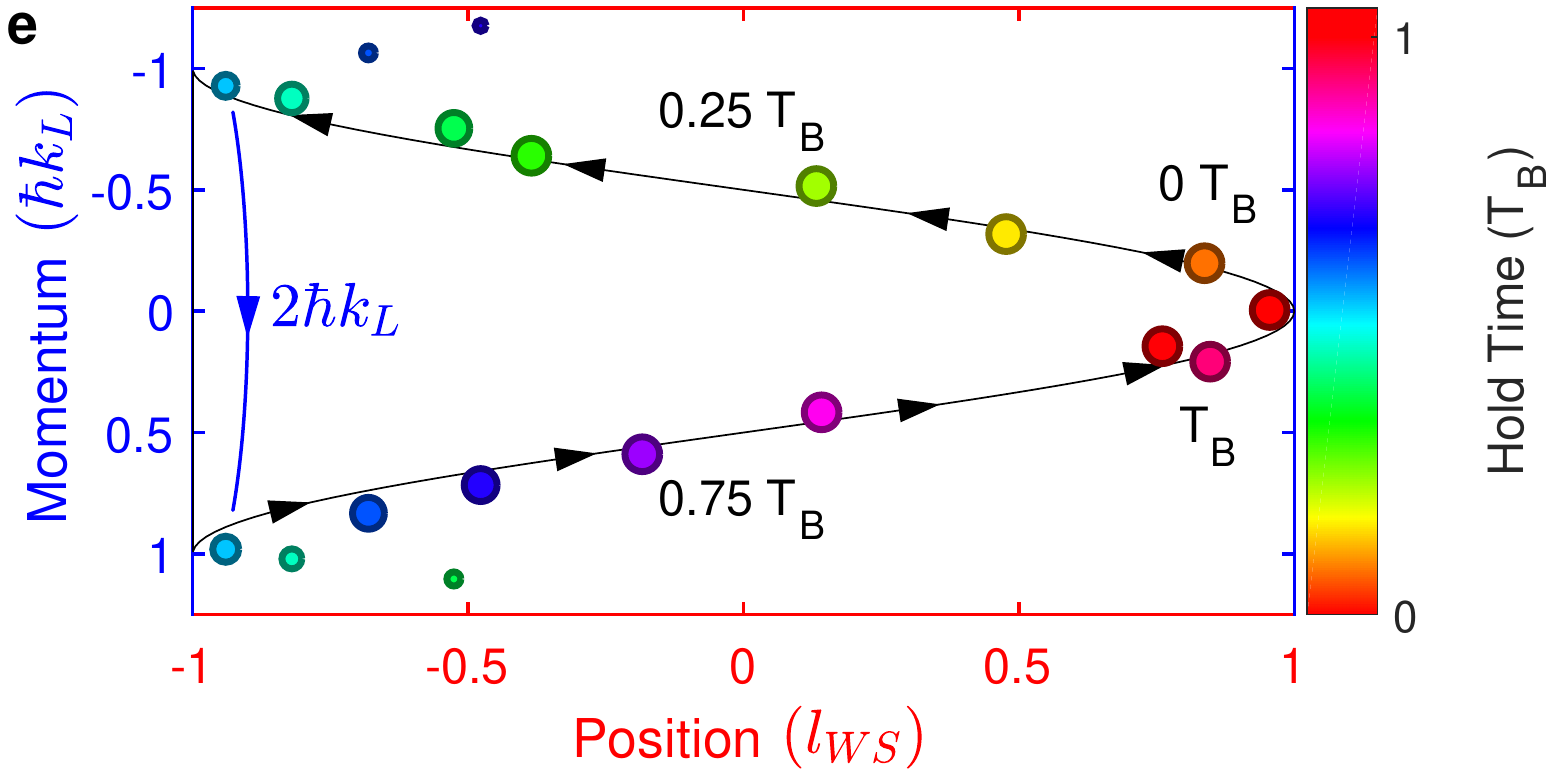}
\caption{Phase-space evolution during a Bloch oscillation. Here $J\!=\! 1549$~Hz and $F\!=\! 23$~Hz. \textbf{(a)}~Absorption images of the atomic distribution after a 3~ms time of flight. The distribution at each hold time is a convolution of initial quasimomentum and position.  \textbf{(b)} Absorption images after varying times of flight, for a hold time of 21~ms. The initial quasimomentum can be extracted via a linear fit. \textbf{(c,d)} Ensemble position and momentum versus time during a Bloch oscillation. Where multiple momentum peaks can be distinguished, both are plotted. \textbf{(e)} Combining (c) and (d) yields the phase space evolution of the ensemble during the first Bloch period $T_B\! =\! 1/f_B$.  Color map corresponds to the hold time in units of $T_B$. Where multiple momentum peaks can be distinguished, both are plotted with symbol size indicating relative atom number. At $0.5 T_B$, Bragg scattering is observed as a discontinuity of $2\hbar k_L$ in the measured momentum distribution.  }
\label{fig:BOcompareTOF}
\end{figure}

\begin{figure}[t!]
\includegraphics[width=\linewidth]{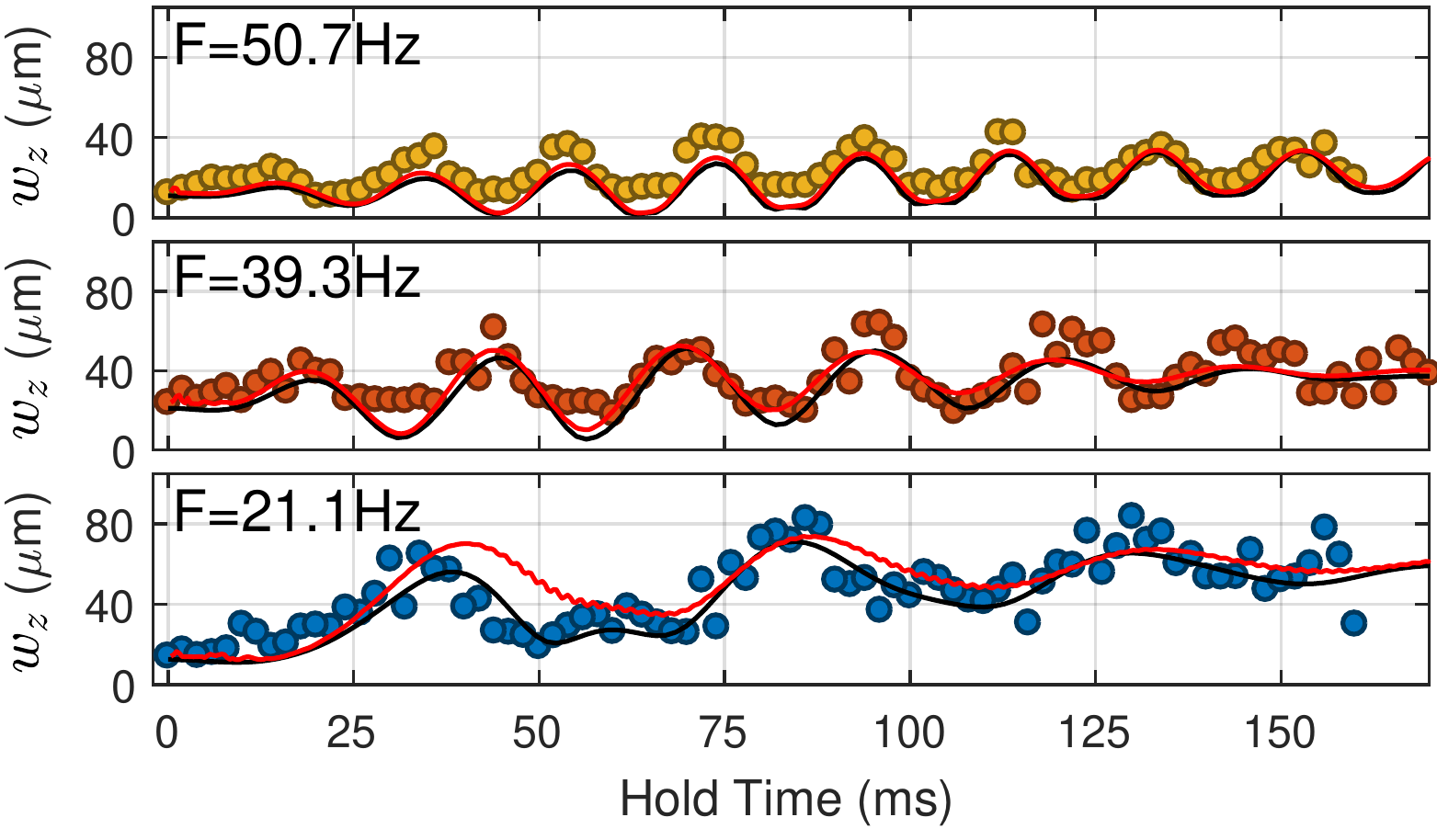}
\caption{Ensemble width evolution during Bloch oscillations. 
Measured time evolution of the cloud width $w_z$ for different forces with $J=1549$~Hz. Black lines correspond to semi-classical evolution, using the calculated band dispersion, of an ensemble with the measured initial width,
and red lines to numerical GPE results. At lower values of $F$ (higher values of $l_{WS}$), the effects of force inhomogeneity are more pronounced.}
\label{fig:breath}
\end{figure}

To elucidate the connection between real-space Bloch oscillations and previously-studied momentum-space Bloch oscillations, we measure the quasimomentum evolution. In these experiments, instead of imaging the in-situ distribution, we ramp down the lattice depth over 100~$\mu$s to map the quasimomentum onto momentum, turn off all traps, and perform absorption imaging of the spatial distribution after some time of flight. For finite time of flight, this image is a convolution of the quasimomentum distribution and the initial spatial distribution.  By performing this measurement at multiple different times of flight we can deconvolve these distributions to measure the evolution of both quasimomentum and position. Fig.~\ref{fig:BOcompareTOF}a shows the image of the atomic ensemble after variable hold times and a constant time of flight of 3~ms. The asymmetrical evolution is a result of the position-momentum convolution. At each hold time we repeat the measurement for various times of flight (Fig.~\ref{fig:BOcompareTOF}b) and fit the linear translation of the ensemble to extract the quasimomentum. This procedure allows direct comparison of the time evolution of position (Fig.~\ref{fig:BOcompareTOF}c) and quasimomentum (Fig.~\ref{fig:BOcompareTOF}d). The measured quasimomentum evolves linearly until it Bragg scatters at the edge of the first Brillouin zone. The turning points of the position-space Bloch oscillations are coincident with the measured Bragg scattering events.  As a demonstration of the unique scientific utility of accessing both real- and momentum-space Bloch dynamics in the same experiment, we plot in Fig.~\ref{fig:BOcompareTOF}e the position-momentum phase space evolution over a single Bloch cycle with period $T_B$. This directly-measured phase-space image of a Bloch oscillation is the second main result we report.

In addition to center-of-mass oscillations, we also observe oscillations in the width of the atomic cloud. In contrast to previous experiments initialized at a single lattice site~\cite{Preiss2015} and previous theoretical predictions for an ensemble~\cite{Gaul2011StabilityNonlinearity,Dominguez-Adame2010BeyondOscillations}, here the width oscillations are dominated by the inhomogeneity in applied force across the large transport distance.  We confirm this experimentally by measuring the width oscillations for different forces at a constant tunneling rate  and comparing them to semi-classical predictions and numerical GPE calculations (see Fig.~\ref{fig:breath}). The good agreement with the two theoretical calculations indicates that the width evolution is a well-understood consequence of force inhomogeneity.

\begin{figure}[t]
\includegraphics[width=\linewidth]{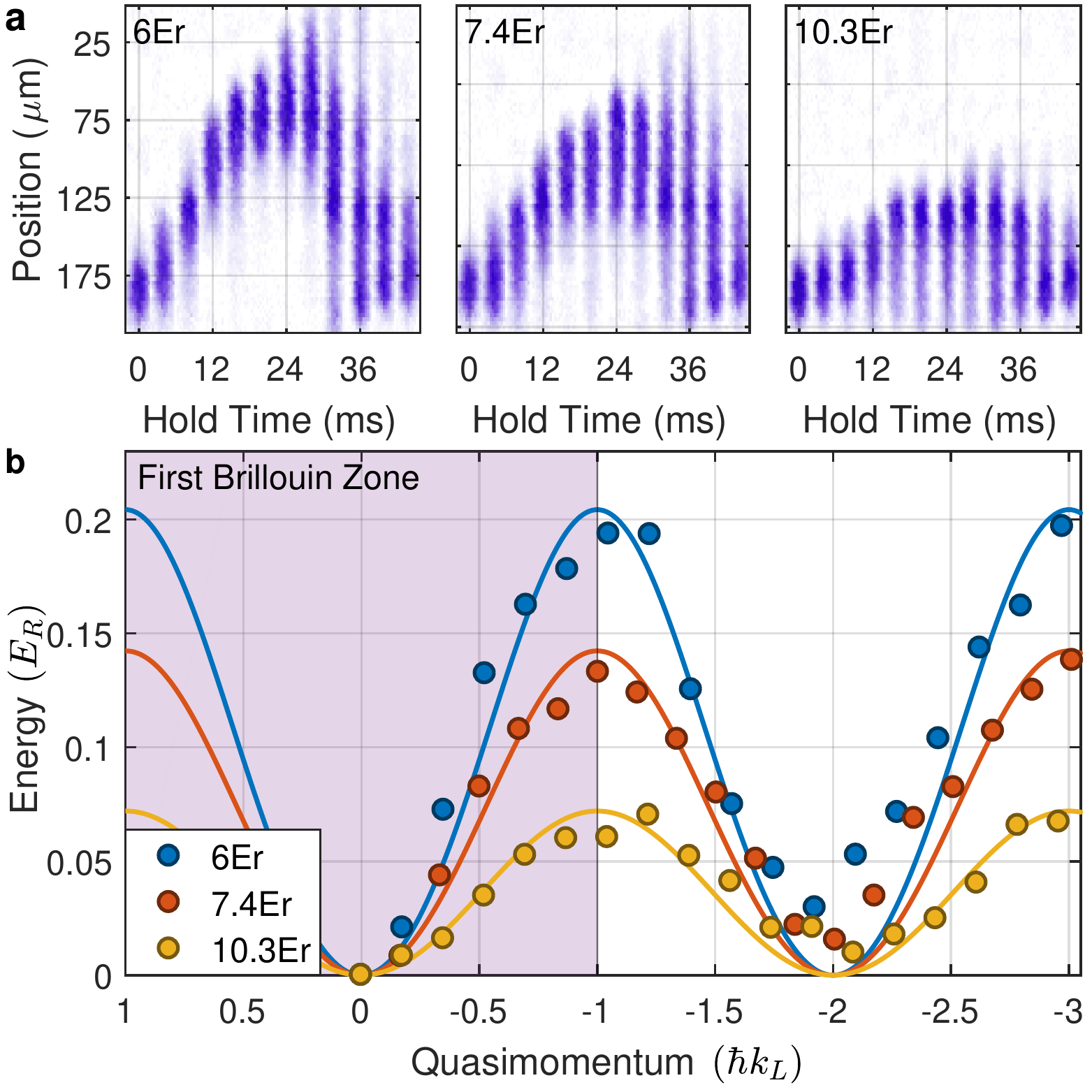}
\caption{Band imaging with position-space Bloch oscillations.  \textbf{(a)} Time-series images of atomic ensembles undergoing Bloch oscillations in lattice depths of 6~$E_R$, 7.4~$E_R$, and 10.3~$E_R$. \textbf{(b)} Measured band structure $E(k)$ for the three lattice depths (circles), determined by scaling the measured center positions from (a) according to Eq.~(\ref{eq:dE}). Lines show independently-calculated band structures with no adjustable parameters.}
\label{fig:BandImaging}
\end{figure}
Strikingly, macroscopic position-space Bloch oscillations can also be used to directly image the lattice band structure.  Eq.~(\ref{eq:com}) indicates that for constant force the band dispersion $E(k)$ is simply the center-of-mass time evolution $x(t)$ during a Bloch cycle with position and time scaled thusly:
\begin{align}
E&=\frac{hf_B}{d} x,\hspace{.3in}  k=\frac{k_L}{2T_B}t. \label{eq:dE}
\end{align}
CCD images of real-space Bloch oscillations like those shown in Fig.~\ref{fig:mixOD} can thus be interpreted as direct images of the lattice band dispersion $E(k)$, in a manner reminiscent of the interpretation of angle-resolved photoemission maps in condensed matter experiments~\cite{duca2015,damascelli2003}.  The Bloch frequency $f_B$ sets the spatial magnification of the band imaging, which can be varied.  For a Bloch frequency of 22~Hz, the calculated energy-position conversion factor is 41.6~Hz/$\mu$m. In Fig.~\ref{fig:BandImaging} we demonstrate this band imaging technique by comparing the measured position-space Bloch oscillations to the theoretically calculated band structure for a variety of lattice depths. We find good agreement with the expected band structure at each lattice depth without the use of any fit parameters. Asymmetries in the measured band structure are consistent with expectations from the measured force inhomogeneity, and represent the current technical limit on this new band imaging technique. The demonstration of direct band imaging with position-space Bloch oscillations is the third main result we report.

\begin{figure}[t]
\includegraphics[width=\linewidth]{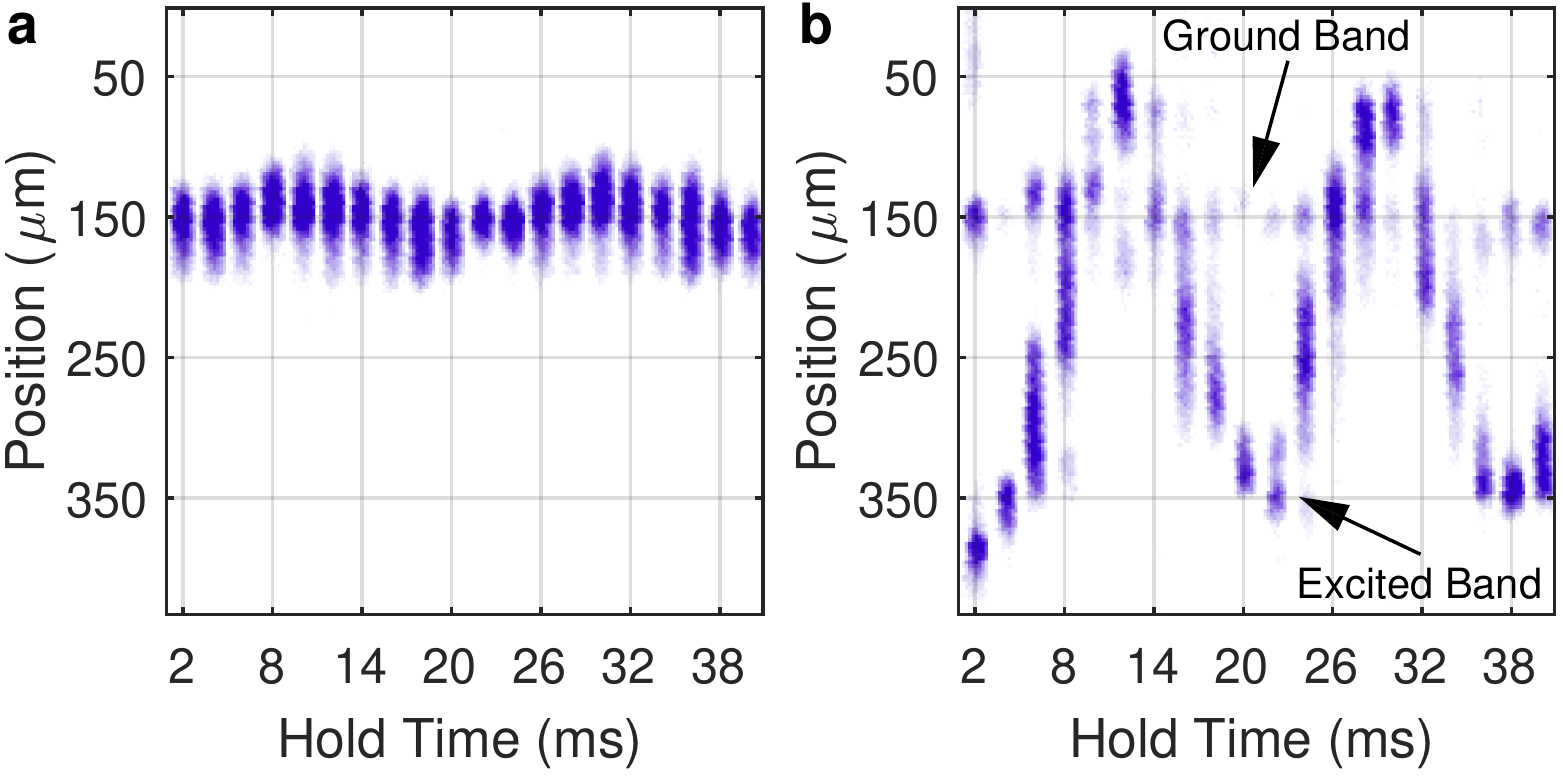}
\caption{Position-space Bloch oscillations in an excited band. The lattice depth is set to 9.5$E_R$ and the Bloch frequency to $50.7$~Hz. {\bf(a)}~Time-series absorption image of ground-band Bloch oscillations.  {\bf (b)} Time-series image of Bloch oscillations in the first excited band, which are characterized by the same frequency as ground-band oscillations, but exhibit a significantly larger amplitude due to the larger bandwidth. Also visible is a remnant fraction of atoms in the ground band, due to imperfect excitation. These Bloch-oscillate independently.}
\label{fig:pband}
\end{figure}

Real-space Bloch oscillations are not limited to the ground band of the lattice.  In excited bands of a sinusoidal lattice, much larger-amplitude position-space Bloch oscillations should be possible due to the increased bandwidth. To explore excited-band real-space Bloch dynamics, we transfer population to the first excited band by applying a resonant amplitude modulation pulse after the atoms have Bloch oscillated to $k=0.5k_L$ in the ground band. The resulting excited-band dynamics are compared to ground-band dynamics in Fig.~\ref{fig:pband}.  Measured Bloch oscillations in the excited band are, as predicted, characterized by a larger amplitude than ground-band oscillations. The amplitude ratio is consistent with theoretical expectations from band structure calculations. 

In addition to their fundamental interest, real-space Bloch oscillations may serve as a useful tool for quantum metrology and spatiotemporal control.  The fine control over transport enabled by the techniques demonstrated here may be of use for atom-interferometric force sensing and gradiometry~\cite{latticeforcemdasurementproposal,BOsforforcesensing,alpha-LKB-BOs,holger-BOsforAI}. New dispersion-control schemes may be enabled by spatially addressing different phases of a Bloch oscillation. Interferometric probes similar to those reported in \cite{Kling2010} but in real space are another possible avenue of exploration. Finally, the simplicity of the full phase-space evolution and band dispersion probing techniques we demonstrate opens up the possibility of imaging dynamics and band structures in more complex systems; particularly interesting possibilities include the study of higher-dimensional lattices, Floquet-hybridized bands, and topologically nontrivial bands~\cite{Aidelsburger2014,Li1094,Fläschner2018}.   

In conclusion, we report the experimental observation of position-space Bloch oscillations in an ultracold gas, in both ground and excited bands.  The dependence of oscillation amplitude on applied force and lattice depth are in good agreement with theory. We have used these real-space Bloch oscillations to directly image the full phase-space evolution during a Bloch oscillation, and to perform direct imaging of the structure of a Bloch band.

The authors thank Peter Dotti and Ethan Simmons for experimental assistance. DW acknowledges support from the Army Research Office (PECASE W911NF1410154 and MURI W911NF1710323). RD and TM thank the PC$^2$ (Paderborn Center for Parallel Computing) for providing computing time. VVK was supported by the FCT (Portugal) under Grant No.~UID/FIS/00618/2013.


\begin{thebibliography}{10}

\bibitem{Bloch1929}
F.~Bloch,
\newblock Zeitschrift f{\"u}r Physik {\bf 52}, 555 (1929).

\bibitem{Zener523}
C.~Zener,
\newblock Proc. R. Soc. A {\bf 145}, 523 (1934).

\bibitem{meier-BOsuperlattice}
J.~Feldmann et~al.,
\newblock Phys. Rev. B {\bf 46}, 7252 (1992).

\bibitem{opticalblochosc}
R.~Morandotti, U.~Peschel, J.~S. Aitchison, H.~S. Eisenberg, and Y.~Silberberg,
\newblock Phys. Rev. Lett. {\bf 83}, 4756 (1999).

\bibitem{dahan1996}
M.~Ben~Dahan, E.~Peik, J.~Reichel, Y.~Castin, and C.~Salomon,
\newblock Phys. Rev. Lett. {\bf 76}, 4508 (1996).

\bibitem{kasevichatomintf}
B.~P. Anderson and M.~A. Kasevich,
\newblock Science {\bf 282}, 1686 (1998).

\bibitem{arimondo-meanfieldBOs}
O.~Morsch, J.~H. M\"uller, M.~Cristiani, D.~Ciampini, and E.~Arimondo,
\newblock Phys. Rev. Lett. {\bf 87}, 140402 (2001).

\bibitem{inguscio-interactionBOs}
M.~Fattori et~al.,
\newblock Phys. Rev. Lett. {\bf 100}, 080405 (2008).

\bibitem{nagerlinteractionBOs}
M.~Gustavsson et~al.,
\newblock Phys. Rev. Lett. {\bf 100}, 080404 (2008).

\bibitem{Gaul2011}
C.~Gaul, E.~D\'{\i}az, R.~P.~A. Lima, F.~Dom\'{\i}nguez-Adame, and C.~A.
  M\"uller,
\newblock Phys. Rev. A {\bf 84}, 053627 (2011).

\bibitem{Meinert2014}
F.~Meinert et~al.,
\newblock Physical Review Letters {\bf 112}, 193003 (2014).

\bibitem{Preiss2015}
P.~M. Preiss et~al.,
\newblock Science {\bf 347}, 1229 (2015).

\bibitem{Haller2010}
E.~Haller et~al.,
\newblock Phys. Rev. Lett. {\bf 104}, 200403 (2010).

\bibitem{Alberti2009}
A.~Alberti, V.~V. Ivanov, G.~M. Tino, and G.~Ferrari,
\newblock Nature Physics {\bf 5}, 547 (2009).

\bibitem{buchleitner2003}
A.~Buchleitner and A.~R. Kolovsky,
\newblock Phys. Rev. Lett. {\bf 91}, 253002 (2003).

\bibitem{VVKgapsolitons}
Y.~V. Bludov, V.~V. Konotop, and M.~Salerno,
\newblock EPL (Europhysics Letters) {\bf 87}, 20004 (2009).

\bibitem{VVK-longBOs}
M.~Salerno, V.~V. Konotop, and Y.~V. Bludov,
\newblock Phys. Rev. Lett. {\bf 101}, 030405 (2008).

\bibitem{Driben-nonlinBOs}
R.~Driben, V.~V. Konotop, T.~Meier, and A.~V. Yulin,
\newblock Scientific Reports {\bf 7}, 3194 (2017).

\bibitem{Pollack2009}
S.~E. Pollack et~al.,
\newblock Physical Review Letters {\bf 102}, 090402 (2009).

\bibitem{Gaul2011StabilityNonlinearity}
C.~Gaul, E.~D{\'{i}}az, R.~P.~A. Lima, F.~Dom{\'{i}}nguez-Adame, and C.~A.
  M{\"{u}}ller,
\newblock Physical Review A - Atomic, Molecular, and Optical Physics {\bf 84},
  1 (2011).

\bibitem{Dominguez-Adame2010BeyondOscillations}
F.~Dom{\'{i}}nguez-Adame,
\newblock European Journal of Physics {\bf 31}, 639 (2010).

\bibitem{duca2015}
L.~Duca et~al.,
\newblock Science {\bf 347}, 288 (2015).

\bibitem{damascelli2003}
A.~Damascelli, Z.~Hussain, and Z.-X. Shen,
\newblock Rev. Mod. Phys. {\bf 75}, 473 (2003).

\bibitem{latticeforcemdasurementproposal}
I.~Carusotto, L.~Pitaevskii, S.~Stringari, G.~Modugno, and M.~Inguscio,
\newblock Phys. Rev. Lett. {\bf 95}, 093202 (2005).

\bibitem{BOsforforcesensing}
G.~Ferrari, N.~Poli, F.~Sorrentino, and G.~M. Tino,
\newblock Phys. Rev. Lett. {\bf 97}, 060402 (2006).

\bibitem{alpha-LKB-BOs}
P.~Clad\'e et~al.,
\newblock Phys. Rev. Lett. {\bf 96}, 033001 (2006).

\bibitem{holger-BOsforAI}
B.~Estey, C.~Yu, H.~M\"uller, P.-C. Kuan, and S.-Y. Lan,
\newblock Phys. Rev. Lett. {\bf 115}, 083002 (2015).

\bibitem{Kling2010}
S.~Kling, T.~Salger, C.~Grossert, and M.~Weitz,
\newblock Phys. Rev. Lett. {\bf 105}, 215301 (2010).

\bibitem{Aidelsburger2014}
M.~Aidelsburger et~al.,
\newblock Nature Physics {\bf 11}, 162 EP  (2014).

\bibitem{Li1094}
T.~Li et~al.,
\newblock Science {\bf 352}, 1094 (2016).

\bibitem{Fläschner2018}
N.~Fl{\"a}schner et~al.,
\newblock Nature Physics {\bf 14}, 265 (2018).

\end{thebibliography}
 \bibliographystyle{aip}

\end{document}